%% file: ms.tex
\documentclass[preprint2]{aastex}

\slugcomment{To appear on ApJ}

\shorttitle{UV nuclei of radio galaxies}
\shortauthors{Chiaberge et al.}

\begin{document}

\title{The  nuclei of  radio  galaxies  in the  UV:  the signature  of
different emission processes\thanks{Based  on observations obtained at
the  Space  Telescope Science  Institute,  which  is  operated by  the
Association of  Universities for Research  in Astronomy, Incorporated,
under NASA contract NAS 5-26555.}}


\author{Marco Chiaberge\altaffilmark{2}, F. Duccio Macchetto\altaffilmark{3}, 
William B. Sparks}
\affil{Space Telescope Science Institute, 3700 San Martin Drive,
Baltimore, MD 21218}
\email{chiab@stsci.edu}

\author{Alessandro Capetti}
\affil{Osservatorio  Astronomico di  Torino,  Strada Osservatorio  20,
I-10025   Pino   Torinese,   Italy}   
\author{ Mark G. Allen}
\affil{Space Telescope Science Institute, 3700 San Martin Drive,
Baltimore, MD 21218}
\and   
\author{Andr\'e   R.   Martel}
\affil{Department of Physics  and Astronomy, Johns Hopkins University,
3400 N. Charles Street, Baltimore, MD 21218}

\altaffiltext{2}{ESA fellow}
\altaffiltext{3}{On assignment from ESA}




\begin{abstract}
We have studied the nuclei of 28 radio galaxies from the 3CR sample in
the  UV band.   Unresolved  nuclei (central  compact  cores, CCC)  are
observed in 10 of the 13 FR~I,  and in 5 of the 15 FR~II.  All sources
that do not  have a CCC in the  optical, do not have a CCC  in the UV.
Two FR~I (3C~270 and 3C~296) have a CCC in the optical but do not show
the  UV counterpart.   Both of  them show  large dusty  disks observed
almost edge-on, possibly  implying that they play a  role in obscuring
the nuclear  emission.  We have measured  optical--UV spectral indices
$\alpha_{\rm o,UV}$  between $\sim 0.6$ and  $\sim 7.0$ ($F_\nu\propto
\nu^{-\alpha}$).  BLRG  have the flattest spectra and  their values of
$\alpha_{\rm o,UV}$ are also confined to a very narrow range.  This is
consistent with radiation produced in a geometrically thin, optically
thick accretion disk.  On the  other hand, FR~I nuclei, which are most
plausibly   originated  by   synchrotron  emission   from   the  inner
relativistic jet, show a wide  range of $\alpha_{\rm o,UV}$.  There is
a  clear  trend  with  orientation  in that  sources  observed  almost
edge--on  or with  clear signs  of dust  absorption have  the steepest
spectra.   These observations imply  that in  FR~I obscuration  can be
present,  but  the  obscuring   material  is  not  in  a  ``standard''
geometrically thick torus.  The most striking difference between these
absorbing structures and the classic AGN ``tori'' resides in the lower
optical depth of the FR~I obscuring material.
\end{abstract}




\section{Introduction}

In  the framework  of  the unification  models  for radio-loud  active
galactic  nuclei, radio galaxies  are believed  to be  the misoriented
counterparts of quasars  and blazars (Barthel 1989; for  a review, see
Urry  \&  Padovani 1995).   Although  observationally  this scheme  is
mainly supported by the comparison of the extended components of these
classes  of AGN,  a  direct study  of  their nuclear  properties is  a
crucial  tool  both  to  confirm   such  a  framework,  and  to  infer
fundamental  information  about   the  innermost  structure  of  these
sources.

In quasars  and blazars, the nuclear components  dominate the observed
radiation.  Their  origin is generally  believed to reside  in thermal
emission  from the accretion  disk and  non-thermal emission  from the
relativistic jet,  respectively.  In  the  case  of blazars,
the jet is observed almost along its axis and its emission is strongly
enhanced by relativistic  beaming \cite{antoulve}.  In radio galaxies,
all of these components should still be present, although the presence
of  obscuring  structures in  the  inner  few  pc, which  is  strictly
required by  the unification models at  least in the case  of the most
powerful  sources  (FR~II),  might  hamper their  direct  observation.
Moreover, the  inner jet emission  should be strongly  de-amplified in
radio galaxies, due to the much larger angle of the line--of--sight to
jet direction.

The  detection of  faint nuclear  optical components  (central compact
cores,  CCC, or  ``nuclei'') in  3CR radio  galaxies, which  are still
unresolved even in HST  images \cite[hereafter Paper~I]{pap1} allow us
to directly  investigate the properties  of the optical  emission from
the active nucleus in this class of sources.

The  picture which  emerges from  these studies  is that  nearby radio
galaxies' nuclei  have two  main flavors.  A  large fraction  of FR~II
appear  to be  consistent  with the  currently  accepted scheme:  they
either  show strong optical  nuclei associated  with narrow  and broad
emission  lines,  or absorbed  nuclei  (either  not  detected or  seen
through scattered  light) in objects  with only narrow  emission lines
\cite[hereafter  Paper~II]{pap4,pap2}.  On the  other hand,  most FR~I
have  unobscured synchrotron-dominated optical  nuclei, low-efficiency
radiating accretion disks  (e.g.  Advection Dominated Accretion Flows,
ADAF,  Rees at  al. 1982,  Narayan \&  Yi 1995)  and  lack substantial
emission lines regions.  Surprisingly, a significant fraction of FR~II
in which broad lines are  not detected, show faint optical nuclei with
optical--radio properties  undistinguishable from those  of FR~I.  The
nature of  these sources is still unclear,  although their consistency
with the  FR~I nuclei  suggests that  at least some  of them  might be
unobscured  synchrotron  nuclei as  well.   These  results have  found
support in the recent observations by Whysong \& Antonucci (2001) of a
strong  $10\mu$  nuclear thermal  component  in  Cygnus~A (3C~405,  an
FR~II), which is not observed in M~87 (3C~274, an optically unobscured
FR~I; see also Perlman et al. 2001).

In light of these results, a classification of radio galaxies based on
their nuclear  properties seems  to be more  closely related  to the
physical process occurring in the central regions of the sources rather
than in  the old FR~I/FR~II morphological dichotomy.   This appears to
be  (at  least  qualitatively)  consistent with  the  dual--population
scheme  proposed   recently  \cite{jacksonwall},  which   unifies  the
different sources mainly on the basis of their spectral properties.

However,  due to  the lack  of  complete spectral  information on  the
nuclei, several  questions are still waiting for  a definitive answer:
are FR~I and  FR~II nuclei intrinsically different?  What  is the role
of obscuration in FR~I?  What is the role of the different ``flavors''
of radio galaxies in the AGN paradigm?

In  this paper,  we  test the  nature  of radio  galaxy nuclei,  using
HST/STIS  UV observations.   High resolution  UV data  are  crucial in
order to test  the new picture of the radio  galaxy dichotomy. In view
of  the relatively low  intensity of  the underlying  stellar emission
from  the  host   galaxy  the  detection  of  CCC   in  UV  images  is
straightforward.  Due to the  reasonably large difference in frequency
between UV and optical data, we  can derive the spectral shape in this
critical region, where different  spectral properties are expected for
different origins of the observed nuclear components.  Furthermore, UV
emission  is very  sensitive  to obscuration  by  dust, therefore  the
presence of  an even moderate amount  of absorption along  the line of
sight  to  the nuclei  will  clearly  affect  their observed  spectral
properties.

The  paper is  organized  as follows:  in  Section \ref{thesample}  we
describe the  sample and the HST observations;  in Section \ref{uvccc}
we outline our  method for the detection and  photometry of the nuclei
in the UV images, and compare it to what has been done in the optical.
In Section \ref{results} we show  our results and we combine them with
the  available  radio  and  optical  nuclear data,  while  in  Section
\ref{discussion} we  discuss the implications  of our results  for the
nuclear  structure  and  the  origin  of  the  emission.   In  Section
\ref{conclusions} we  present a  summary of our  findings and  we draw
conclusions.

\section{The sample and the HST observations}
\label{thesample}

We analyze a sample of  nearby ($z<0.1$) FR~I and FR~II radio galaxies
belonging  to the  3CR  catalog, for  which  both optical  and UV  HST
observations are  available.  The sample comprises  28 radio galaxies,
13 of them morphologically classified as  FR~I and 15 as FR~II. Of the
15  FR~II,  7 are  HEG  (high excitation  galaxies),  5  are LEG  (low
excitation galaxies;  Hine \& Longair 1979, Jackson  \& Rawlings 1997)
and 3 are broad--line radio  galaxies (BLRG).  The list of the sources
is shown in Table \ref{tab1}.

\input tab1.tex

All  objects, except  for  3C~78, 3C~264  and M~87\footnote{3C~78  and
3C~264 have  been observed  with STIS as  part of program  8233, while
M~87 have  been observed  as part of  program 8140, P.I.  J. Biretta},
have been observed with the HST as part of the STIS UV snapshot survey
of 3CR radio sources (Allen et  al.  2002).  We have excluded from the
sample 3C~231  (M~82) which is a starburst  galaxy.  STIS observations
have  been made  with the  NUV--MAMA  detector and  filters with  peak
sensitivity at  $\sim 2300$  \AA.  In particular,  most of  the images
were obtained with the F25SRF2 long pass filter, which excludes strong
contamination from  the Ly$\alpha$  emission line within  the redshift
range of our  sources.  The brightest objects have  been observed with
the narrower band F25CN182 filter, in order to avoid saturation. M~87,
and 3C~78 have also been observed by STIS NUV--MAMA, using the F25QTZ
filter (whose characteristics are similar to F25SRF2) while 3C~264 has
been observed using the F25CN182 filter.  

The log of optical and UV observations is presented in Table~\ref{log}
where it can  be seen that for 4  sources (M~87, 3C 78, 3C  264 and 3C
317) the      observations       occurred      simultaneously      (or
near-simultaneously). Given that all  sources in the sample are likely
to be  intrinsically variable, the time lapse  between observations of
the other  24 sources might be  significant and we  discuss this point
further in section \ref{variability}.

\input tab2.tex

The sample is  not statistically complete, because a  few sources were
not observed during the SNAPSHOT campaign.  However, all the different
spectral and  morphological types present  in 3CR radio  galaxies with
$z<0.1$ are well represented.

\section{Detection and photometry of UV CCC}
\label{uvccc}

As shown  in detail by  Allen et al. (2002) the morphology  of radio
galaxies as observed  in the UV band can  be significantly different
from what is seen in the  optical images.  This certainly holds in the
nuclear region, where dust features dramatically absorb the underlying
emission  in  the UV  and  clumps  of  (likely) star  formation  are
present. Also, in that band  the stellar emission from the host galaxy
is substantially reduced with respect to optical (R-band) frequencies,
typically   by   a   factor   of   $\sim  15$,   in   an   $F_\lambda$
representation.

Most  of  the UV  images  have very  low  level  of stellar  emission.
Therefore,  while the search  for optical  CCC has  been based  on the
analysis of the surface brightness  profiles of the central regions of
the galaxies  (Papers I  \& II),  here we search  directly for  the UV
counterparts  of the  optical CCC.   In addition,  we also  search for
unresolved UV CCC in galaxies which  do not have optical CCC, but find
none.  All of the UV CCC sources have FWHM $\sim 0.05^{\prime\prime} -
0.07^{\prime\prime}$, indicating that they are unresolved by HST.

Of the 13 FR~I galaxies in our sample, 10 (or 77\% of the sample) have
a  UV nucleus.   In the  optical, 12  (92\%) have  a CCC.   Whenever a
galaxy has a nucleus in the UV,  it also has a nucleus in the optical.
Of all the FR~Is in the sample, only 3C~305 does not have a CCC either
in the  optical or  in the UV.   Interestingly, the two  sources which
lack  the UV CCC  but do  have it  in the  optical (namely  3C~270 and
3C~296)  have prominent nuclear  dusty disks  observed almost  edge on
\cite{cc0,martel}.   These  disks  are  clearly visible  in  both  the
optical and the UV images.

Of the 15 FR~II galaxies of  our sample, only 5 sources have a nucleus
in the UV. All of them also  have a CCC in the optical. In particular,
a CCC is present in all of the  3 BLRG, and only in $1/5$ and $1/7$ of
the HEG and LEG, respectively.   The BLRG have the brightest nuclei of
the sample.   HST data are  available for 3C~285  both in the R  and V
bands.  A faint CCC is present in the R band image, but this component
is not  detected in the  V band.  In  this case, obscuration  might be
provided  by  a prominent  large  scale  dust  lane, similar  to  that
observed in Centaurus A. In  this latter object, the IR-bright nucleus
vanishes  in  HST images  for  wavelengths  shorter  than $\sim  5000$
\AA~\cite{marconi}.

We have performed aperture photometry of all the nuclei using the IRAF
RADPROF  task, setting  the background  level at  a distance  of $\sim
0.17$  arcsec (7  pixels in  the  STIS/MAMA images)  from the  center.
Counts   were  converted   to   fluxes  adopting   the  HST   internal
calibration\footnote{The  PHOTFLAM  parameter   in  the  image  header
(inverse sensitivity) is defined assuming a flat spectral distribution
in $F_\lambda$ (in units of erg cm$^{-2}$ s$^{-1}$ \AA$^{-1}$).  As we
will show  in the following,  most of the nuclei  have significatively
sloped spectral indices.  We have therefore recalculated the effective
observing wavelength and the value  of the inverse sensitivity with an
iterative process for  all of the sources, using  the SYNPHOT package.
Only  in the case  of 3C~449,  the source  with the  steepest observed
spectrum,  this   significatively  affects  the   effective  observing
wavelength and the  estimate of the flux.}, which  is accurate to 5\%.
However, except for those sources in which the CCC is clearly the only
observed  component in the  UV, the  dominant source  of error  is the
presence  of  some  extended  emission  from the  host  galaxy  and/or
absorption  features  which   might  strongly  affect  the  background
determination.   This  results in  a  typical  error  of $\sim  10\%$,
therefore comparable to that in the optical.

For 3C~270 we  derive a rough upper limit to the  nuclear flux, as the
presence of a dusty disk and emission features in both the optical and
UV images allows us to  identify the position of the nucleus, although
it is  not visible in the  UV.  For the remaining  objects no reliable
upper limits  can be derived.  Also note that  the LEG 3C~35 has  a UV
central component which is not a point source.

\input tab3.tex

\input tab4.tex

\section{Results}
\label{results}

The  results of  the photometry  of the  CCC are  summarized  in Table
\ref{cccfluxes}.   For  completeness,  we  also  include  the  optical
fluxes,  as taken  from Paper~I  and II.   Since in  the UV  band dust
absorption  plays  an  important  role  in  drastically  reducing  the
observed flux, we have de-reddened  the fluxes taking into account the
galactic absorption and adopting the Cardelli et al. (1988) extinction
law (Table \ref{ccc2}).  In the following sections and figures we will
use  the de-reddened  fluxes,  although galactic  extinction does  not
significantly affect  the results  in any of  the sources\footnote{The
maximum  value of  the  color  excess due  to  galactic absorption  is
$E(B-V)=0.167$ in the  case of 3C~449. This value  of $E(B-V)$ affects
the  UV   and  optical   fluxes  by  factors   of  only  3   and  1.4,
respectively.}.

To study  the relationship between the  optical and UV CCC  for the 15
sources of Table  4, we plot their UV flux  and luminosity versus their
optical values.  The same  behavior is observed  in both the  flux and
luminosity regimes:  we see that the UV  CCC span a range  of $\sim 5$
dex in  flux and $\sim 6$  dex in luminosity, whereas  the optical CCC
span smaller ranges (about 1 dex in both flux and luminosity).

\begin{figure}
\plotone{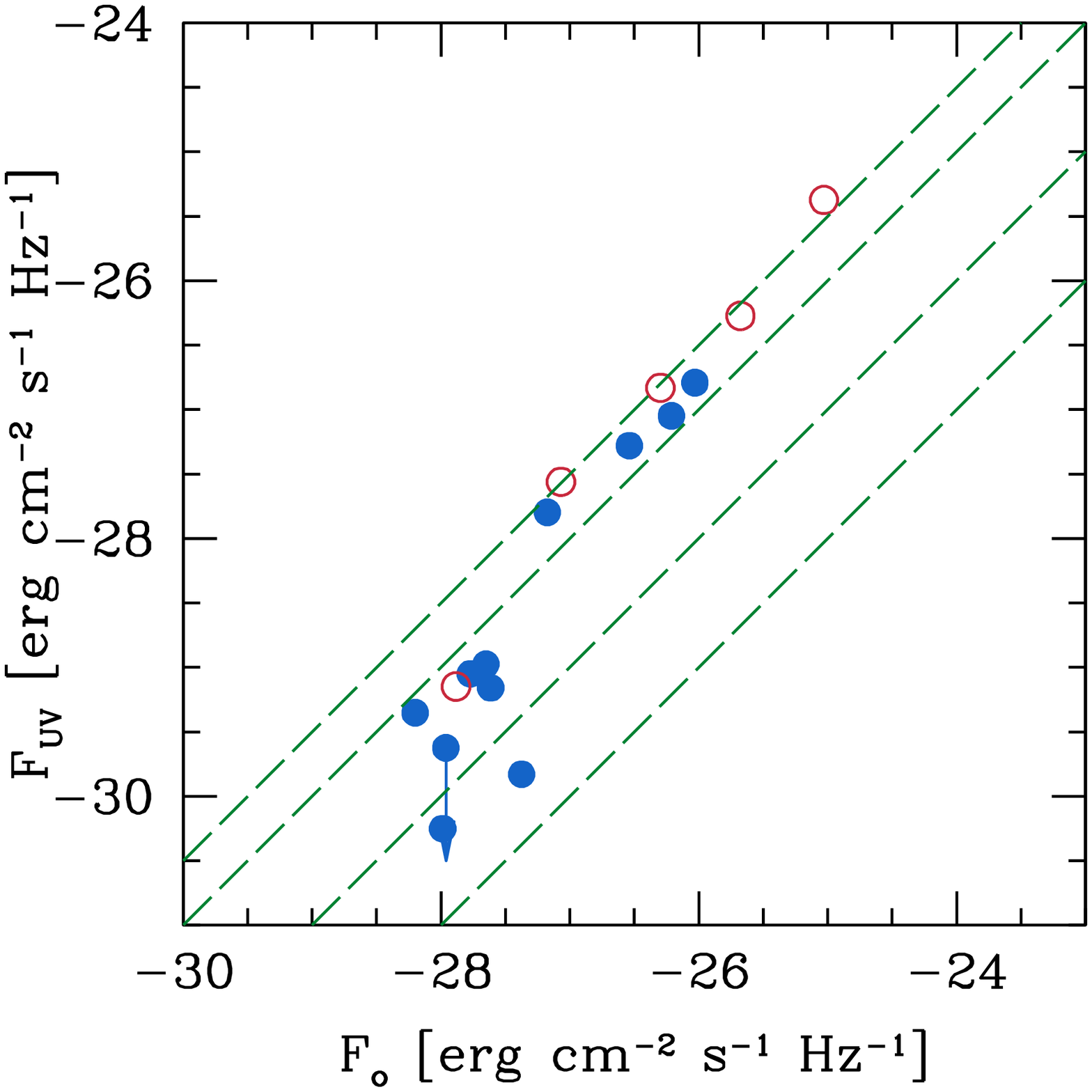}
\caption{UV flux vs. optical flux of the central compact cores. Filled
circles are  FR~I, while FR~II  are represented as empty  circles. The
dashed lines  represent values of $\alpha_{\rm  o,UV}$=1,2,4,6 (top to
bottom).
\label{fofouv}}
\end{figure}

\begin{figure}
\plotone{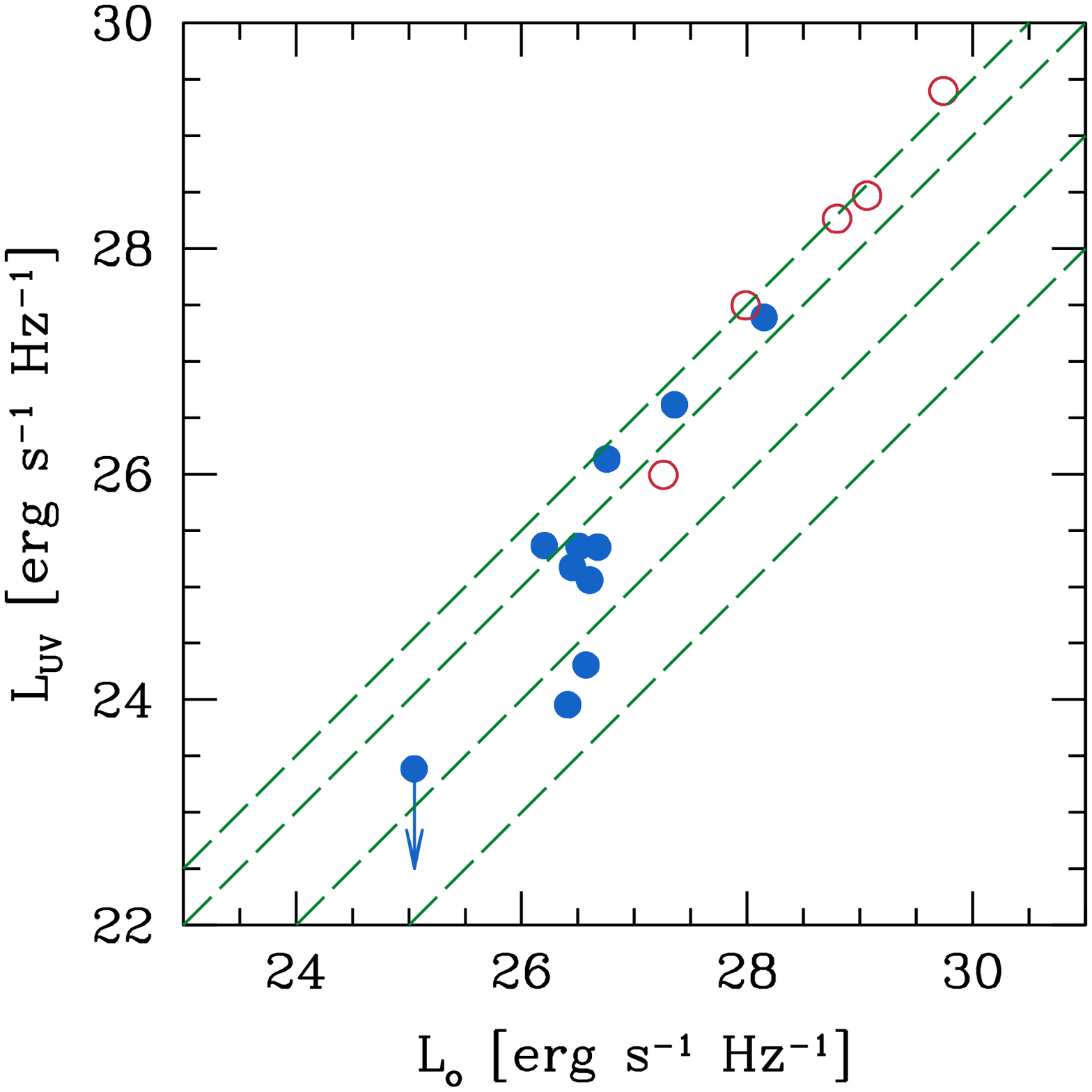}
\caption{UV   vs.   optical   luminosity   of  the   central   compact
cores. Filled circles  are FR~I, while FR~II are  represented as empty
circles. The dashed  lines represent values of $\alpha_{\rm o,UV}$=1,2,4,6
(top to bottom).\label{lolouv}}
\end{figure}

Given the  large difference in frequency between  the optical (R-band)
and UV  observations, we could  derive the broadband  spectral indices
[optical--UV]  with  considerable  greater  accuracy  than  previously
possible (Paper~I).  The values of $\alpha_{\rm o,UV}$, where $\alpha$
is  defined as $F_\nu  \propto \nu^{-\alpha}$,  are reported  in Table
\ref{ccc2}. Typical errors for $\alpha$ are of the order of $\pm 0.1$.
For  reference,   in  Fig   \ref{fofouv}  and  \ref{lolouv}   we  have
overplotted  dashed   lines  corresponding  to   different  optical-UV
spectral indices $\alpha_{\rm o,UV}$=1,2,4,6 (top to bottom).  We find
that the UV and optical  luminosities of the brightest sources sources
are almost  identical.  However, the lower  optical luminosity sources
have much weaker UV emission (by a factor of $\sim 100$).

\begin{figure}[h]
\plotone{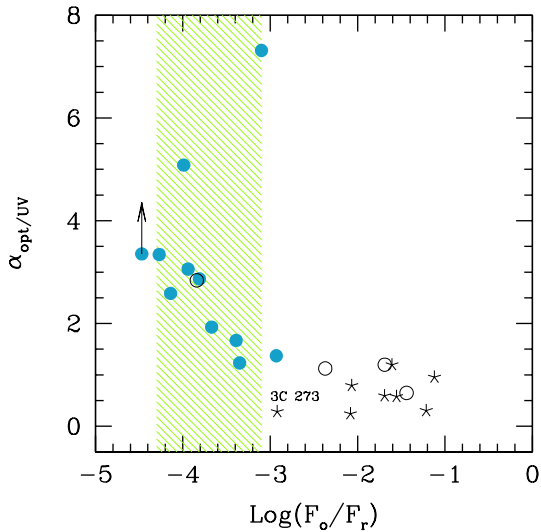}
\caption{$\alpha_{\rm o,UV}$ versus the logarithm of the ratio between
optical CCC  flux to radio core  flux. Filled circles  are FR~I, while
FR~II are  represented as  empty circles.  Stars  represent radio-loud
quasars from the Elvis et al. 1994 sample.  The shaded region
corresponds  to   the  1$\sigma$  dispersion   of  the  radio--optical
correlation found for FR~I nuclei.\label{alrro}}
\end{figure}

In  Fig. \ref{alrro}  we plot  the optical-UV  spectral index  vs. the
ratio  between  the  optical  CCC   and  the  radio  core  flux.   The
quasi-linearity  of  the  FR~I  radio-optical  correlation  implies  a
quasi-constant ratio  of the optical  to radio flux  (corresponding to
$\log (F_o/F_r)  \sim -3.6$),  therefore objects on  the left  side of
this  plot are on  the FR~I  correlation, while  for higher  values of
$\log  (F_o/F_r)$  an optical  excess  is  present.   The shaded  area
corresponds  to the  rms of  the correlation  (see figure  caption for
details). Unfortunately,  no radio core measurements  are available in
the literature  for 3C~198,  therefore its position  is this  plane is
undetermined.   Sources clearly  separate in  this  plane. Broad-lined
FR~II  have  similar  (and  rather  flat)  $\alpha_{\rm  o,UV}$.   The
synchrotron--dominated FR~I have lower values of $F_o/F_r$ by a factor
of ($\sim 10-100$), they lie  in the region of the correlation between
the radio  and optical cores, but  span a large  range in $\alpha_{\rm
o,UV}$.   Interestingly,  the  only  LEG  FR~II with  a  detected  CCC
(3C~388) lies in the region  typical of FR~I.  For comparison, we have
also plotted  radio-loud QSO\footnote{The radio-loud  QSO with $z<0.3$
from the Elvis et al.  (1994) sample are: 3C~206 ($z=0.197$), 3C~323.1
($z=0.264$),  MC2~1635+119 ($z=0.146$),  PHL~1657($z=0.2$) PG~0007+106
($z=0.089$), B2~1028+313 ($z=0.178$),  3c~273 ($z=0.158$) and 4C~34.47
($z=0.206$).}  with $z<0.3$  from the sample of Elvis  et al.  (1994).
Since these sources are at a  higher redshift than our 3C galaxies, we
have used  the I-band  magnitudes and the  UV fluxes measured  at 3000
\AA~  (as taken  from Elvis  et al.   1994) to  mimic  the rest--frame
$\alpha_{\rm o,UV}$.  BLRG  in our sample and the  QSO occupy the same
region  of the plane,  except for  3C~273.  In  the following  we will
discuss the implications  of the different position of  the sources in
this  plane, strongly  supporting a  different origin  of  the nuclear
emission in the various classes of radio galaxies.

\section{Discussion}
\label{discussion}

UV data are a fundamental tool  to study the effects of obscuration in
the central regions  of these galaxies, as this  spectral band is very
sensitive  to  the  presence  of dusty  structures.   Also,  different
emission  processes  for the  nuclei  imply  different  values of  the
optical--UV  spectral  slopes.  By  combining  the  UV  data with  the
already  available  radio  and  optical  data,  we  can  address  some
important questions such as:

$\bullet$ are FR~I and FR~II intrinsically different?

$\bullet$ what  do the  UV observations tell  us about  their physical
nature?

$\bullet$ what is the role of these sources in the AGN paradigm?

$\bullet$ what is the role of obscuration in FR~I?

Qualitatively, FR~II nuclei are brighter and have flatter spectral slopes
than  FR~I.   Such  a behavior  can  be  accounted  for by  two  basic
scenarios: (i) an  intrinsic spectral difference, reflecting different
physical properties,  and (ii) an external reason,  namely a different
amount of  absorption in the  various sources.  Obviously, it  is also
possible that a combination of these two factors contributes to define
the  observed  properties.   Whatever   the  nature  of  the  CCC,  an
increasing column  density naturally both steepens  the spectral slope
and  lowers the  amount of  observed photons.   However,  the physical
processes responsible for the emission  cannot be constrained on the basis
of only the optical and UV fluxes and luminosities.

In the  following sections we will  further discuss the  nature of the
CCC by taking advantage of the already known behavior of the different
sources in the radio-optical plane.

\subsection{Disk dominated vs. synchrotron dominated sources?}

It is   widely  believed that  in  radio--loud  AGN two main  emitting
components play a role in defining  the optical-UV spectral properties
of the nuclear continuum: the thermal emission from the accretion disk
and  the non-thermal synchrotron  radiation from the relativistic jet.
In addition, we  must  consider the role  of  absorption, which  might
significatively  alter  the observed spectra.   Theoretical models for
thermal  disk emission (in   the case of a standard  Shakura--Sunyaev,
optically thick  and geometrically   thin  disk) predict that   such a
component should peak in the UV band, at wavelengths shorter than 1000
\AA,  therefore hard UV spectral  slopes  ($\alpha_{UV} \sim 0.3$) are
expected.  Flat spectral components  with $\alpha_{\rm o,UV} <  1$ are
indeed commonly observed in both  radio--quiet and radio loud QSO (see
e.g.  Elvis et al. 1994).

On the other hand, the spectral slope of synchrotron dominated sources
is poorly constrained, as it depends on physical parameters (mainly on
the electron distribution, the  magnetic field and the beaming factor)
which cannot be  easily determined ``a priori''.  However,  we can try
to estimate it by analogy  with other synchrotron emitting sources for
which the SED is well known.   In the framework of the AGN unification
schemes,  radio  galaxies are  believed  to  constitute the  so-called
parent  population  of  blazars,  therefore the  comparison  with  the
observed properties  of these sources  might be helpful.   Blazars are
objects  in  which   the  jet  is  believed  to   be  observed  almost
``on-axis''.  Their  observed SED (in  a $\log \nu -  \log(\nu F_\nu)$
representation) is composed of two  broad peaks: the lower energy peak
is commonly interpreted as due to synchrotron emission, while the high
energy peak is ascribed to inverse Compton emission.  The frequency of
the peaks can vary substantially from one source to another: the lower
energy peak  can be located between  the IR and the  X-ray band, while
the higher energy peak is  generally in the gamma-ray band. Fossati et
al. (1998) have shown that  the position of the synchrotron peak, which
is well constrained by the  observations, is related to the bolometric
luminosity  of the source.   The lower  the power  of the  object, the
higher the synchrotron peak frequency $\nu_{\rm peak}$.

Clearly,  the  relative  position   of  the  peak  and  the  observing
frequencies strongly affects  the observed UV--optical spectral index.
For objects  in which $\nu_{\rm  peak}$ is significatively  lower than
the observing  range, steep ($\alpha  \sim 1-3$) spectral  indices are
measured, while flatter spectra  ($\alpha \sim 0.5-1$) are observed if
$\nu_{\rm peak}$ is placed at higher frequencies.

Relativistic effects must also be taken into consideration in order to
estimate  the  putative  position  of  the synchrotron  peak  for  our
nuclei.  Since the  emission  from  the jet  is  strongly affected  by
beaming,  the peak  frequency shifts  towards lower  values  and (most
importantly) towards  lower luminosities as the angle  between the jet
axis  and  the line  of  sight increases.   However,  as  a result  of
beaming, it  is possible  that in highly  misoriented objects  such as
most radio  galaxies, the radiation  emitted by a slower  component of
the jet (e.g.  a layer) might dominate the  observed nuclear emission,
while the (faster) component which dominates blazar emission might not
be observed  \cite{pap3}.  Unfortunately,  the spectral
properties of this component are still very poorly known.

The position  of the  objects in the  plane formed by  the optical--UV
spectral index vs the logarithm  of the ratio between optical CCC flux
to radio core  flux, which is shown in Fig.   \ref{alrro} and has been
described in Sect. \ref{results}, can be interpreted as related to the
different  nature of their  nuclear emission.   Objects in  our sample
which  show  an  optical  excess  with respect  to  the  radio-optical
correlation  are   BLRG  FR~II  and  have   flatter  spectral  indices
($\alpha_{\rm  o,UV}\sim  0.6-1.2$).  These  sources  occupy the  same
region of the plane as radio-loud quasars of comparable redshift, from
the \cite{elvis} sample, which are plotted as stars.  The
SED of the QSOs  is well known and it clearly shows  the presence of a
``blue  bump'', which is  commonly interpreted  as the  most prominent
signature  of thermal  emission  from the  accretion  disk.  The  only
quasar  with a lower  value of  $\log (F_o/F_r)$  is 3C~273,  which is
classified as  a blazar.  The  contribution from the  relativistic jet
emission (which  generally shows  strong variations) in  this spectral
region can  indeed alter substantially  the observed spectrum  in this
source \cite{gg}.  In  fact, it lies close to the  region of the FR~I
correlation,  indicating that  thermal  disk emission  is  low in  the
optical band, but  shows a flat $\alpha_{\rm o,UV}  < 1$, implying the
presence of a  strong component at higher energies,  similarly to what
is observed in the other QSO.

We conclude  that our  BLRG  are  compatible with being  thermal--disk
dominated objects.  Their distance  from the shaded region is possibly
determined by the relative contribution of  the disk and jet emission,
which might also be related to the source orientation.

As  pointed out  above,  FR~I  show a  different  behavior. They  have
similar  values of  $F_o/F_r$, since  they lie  on  the radio-optical
correlation, while  they span a  large range in $\alpha_{\rm  o, UV}$,
from $\sim 1$ to the extreme case of 3C~449, for which $\alpha_{\rm o,
UV}=7$. However, taking into  account the above considerations, we can
conclude that slopes significatively flatter or steeper than unity are
not expected  for our sources.   Steep spectral slopes  in synchrotron
spectral components can be observed  only for a relatively small range
of  frequencies, which is  confined to  be well  above the  low energy
peak, but  still before the rising of  the inverse--Compton component.
Therefore, intrinsic  values of $\alpha_{\rm o,UV}$  larger than $\sim
3$ appear to  be inplausible for synchrotron emitting  sources. In the
following section  we show  that a moderate  amount of  absorption can
account for the observed behavior.

\begin{figure}[h]
\plotone{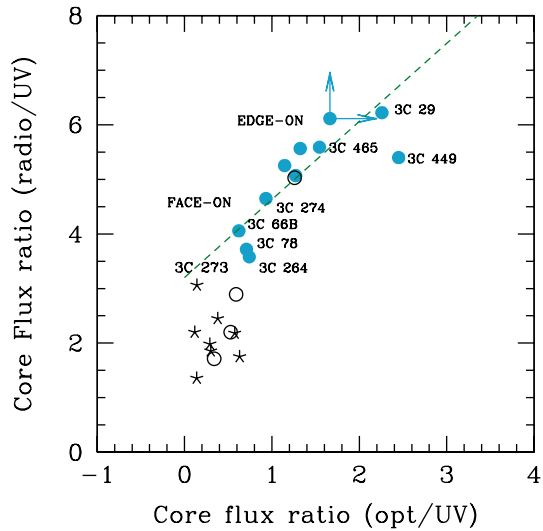}
\caption{The ratio of  the radio core to UV CCC flux  vs. the ratio of
the optical CCC to UV CCC  flux.  Filled circles are FR~I, while FR~II
are represented  as empty circles. Stars  represent radio-loud quasars
from the  Elvis et  al. 1994 sample.   The dashed line  is the
absorption  trail: less absorbed  sources (seen  face--on) are  on the
bottom-left of the plot (section \ref{absorption}).
\label{ruvouv}}
\end{figure}

\subsection{Evidence for a moderate amount of obscuration in FR~I}
\label{absorption}

The apparent  inconsistency between  the synchrotron scenario  and the
presence  of such steep  spectral slopes,  can be  solved if  a modest
amount of  absorption is present, since it  might naturally contribute
to steepen the observed spectrum.   In order to test whether the range
of spectral indices and their  relative amounts of radio, optical, and
UV  emission is  compatible  with  being due  to  absorption, we  have
plotted in Fig. \ref{ruvouv} the ratio of radio to UV emission, versus
the  ratio between optical  and UV  emission. The  dashed line  is the
``absorption trail'', which has been calculated by taking into account
the  effects of  an increasing  amount of  absorption on  the observed
optical and UV fluxes.

All FR~I  (except for  3C~449) are  well aligned to  such a  line. The
FR~II 3C~388  lies on the  absorption trail as  well. If we  assume an
intrinsic slope of  1 for all of these sources,  the range of observed
$\alpha_{\rm o,UV}$ corresponds to  a very small amount of absorption.
A maximum $A_V$ of $\sim 6$ is obtained in the extreme case of 3C~449,
while  all other  sources are  between 0.15  and 3  (the  median being
$A_V=1.3$).   Although  this  is  a  {\it clear  evidence  of  nuclear
absorption  in  FR~I  sources},   these  observations  show  that  the
properties  of  the  absorbing  material  are not  compatible  with  a
standard geometrically thick torus structure. In fact, the position of
the sources along the absorption trail in Fig.\ref{ruvouv} is strictly
connected with orientation.  On the one hand, the sources which appear
to  be the  less absorbed  (namely 3C~264,  3C~66B, 3C~78  and 3C~274,
lying on  the bottom-left of  the plane, close  to the BLRG),  are all
objects in which optical jets are seen.  The presence of such features
has been  interpreted as due  to their relatively small  viewing angle
\cite{sparks} which, as a  result of relativistic beaming, contributes
to enhance the jet radiation.  On the other hand, dusty disks observed
almost edge-on are seen among the most absorbed sources (e.g.  3C~449,
3C~465; Capetti  \& Celotti 1999, Martel  et al.  2000).   In the most
extreme cases  (3C~270, 3C~296) the nuclear source  is clearly visible
in the  optical, but  it is not  present in  the UV.  Such  trend with
orientation implies that the  absorbing material cannot be distributed
either as  a spherical  structure (all sources  should be  affected by
absorption) nor  as a thin disk  where the absorbing  material is well
confined.   In the  latter case,  in fact,  no trend  with orientation
should  be seen:  sources should  be either  absorbed or  not absorbed
without any  intermediate case. However, the  most striking difference
between these structures  and the classic AGN ``tori''  resides in the
lower optical depth  of the FR~I absorbers, which  allow us to observe
the CCC even in the UV in  all FR~I, except for the most extreme cases
of 3C~270 and 3C~296.

In light of  these results, we can confirm the  claim that FR~I nuclei
are generally unabsorbed.  The  moderate amount of absorption observed
in these objects  cannot be ascribed to the  presence of a ``classic''
torus--like  structure which  is typical  of other  AGN, and  might be
instead  accounted  for either  by  extended  (kpc--scale) dust  lanes
(e.g. in the case  of 3C~29, in which the signature of  a dust lane is
clearly  visible in  the  HST images, Sparks et al. 2000),  or by  the
($\sim 100$ pc--scale) dusty disks.  In the latter case, the absorbing
column density must vary smoothly for different viewing angles.

As already  noted above, one FR~II  (3C~388, which is  classified as a
LEG) lies in the region typical of FR~I.  This appears to confirm what
has been already found in Paper~II: 3C~388 belongs to a third class of
FR~II sources whose optical nuclei are indistinguishable from those of
FR~I (we have  called such class ``FR~I--like'').  The  fact that this
object is both on the  radio-optical correlation and does not show any
UV excess,  typical of thermal  disk emission, further  constrains its
origin as due to non-thermal  synchrotron radiation, as it is for FR~I
nuclei.  This also rules out  the possibility that its CCC emission is
produced by a compact scattering  region, reflecting the presence of a
hidden quasar in its nucleus,  since quite flat values of $\alpha_{\rm
o,UV}$ should be observed in  that case. The synchrotron origin of its
nuclear emission, together  with the lack of strong  emission lines in
its spectrum, lead us to  identify 3C~388 with a ``parent source'' for
BL  Lac objects,  in  agreement with  the  ``dual population  scheme''
\cite{jacksonwall}.

\subsection{Variability}
\label{variability}

Spectral variability  is a  common characteristic of  both synchrotron
and   disk--dominated   sources.    In   particular,   highly   beamed
non--thermal sources  such as BL Lac objects  show dramatic variations
on  all timescales (from  minutes to  years).  Since  the relativistic
beaming  factor  is  expected  to  be  significantly  lower  in  radio
galaxies, such behavior is expected to be present, although on longer
timescales.  Clearly, this might  strongly affect the determination of
the  spectral  slope.   Therefore,  it  is  important  to  use,  where
available, simultaneous observations.

Only  three  sources  in  our  sample  (3C~78,  3C~264,  3C~317)  have
simultaneous (within few hours) optical--UV data, while for 3C~274 the
time lag  between optical and UV  observations is only  $\sim 6$ days.
In addition to  the FOC, 3C~317 has also been observed  as part of the
STIS  UV   SNAPSHOT  program\footnote{3C~317  was   observed  by  STIS
NUV-MAMA, with the F25SRF2  filter, on 1999 July 27.}.  Interestingly,
we have found that the flux of the CCC in the STIS observation is $5.1
\times 10^{-17}$ erg cm$^{-2}$  s$^{-1}$ \AA$^{-1}$, a factor of $\sim
10$  brighter  than  in  the  FOC  observation  of  1994.   Such  high
variability   factor  indeed   strongly   supports  the   non--thermal
synchrotron  scenario.   Variability has  been  also  detected in  the
optical   band  in   3C~274,  although   with  much   lower  intensity
\cite{tsvetanov}.

The  determination of  the spectral  slopes  of objects  for which  no
simultaneous  data  are available  might  be  significantly affected  by
variability.  However,  we are confident that the  general behavior of
the CCC is correctly represented  by our estimates.  For the BLRG this
is mainly supported by the fact that all of them show similar spectral
indices.   For the  FR~I, both  their position  in the  plane  of Fig.
\ref{ruvouv} and  the close  connection between steep  $\alpha_{\rm o,
UV}$ and  the presence of  obscuring structures are strong  clues that
obscuration and not variability is indeed the main physical reason for
the observed steep slopes.

We  conclude that  the general  behavior  is well  represented by  the
picture outlined  above, although we cannot definitively  rule out the
possible presence of variability in individual sources.

\section{Summary and conclusions}
\label{conclusions}

We have analyzed images of 28  nearby 3C radio galaxies for which both
optical  and UV  HST images  are available.   We have  found  that all
objects which show  an UV CCC, also show it in  the optical.  Only two
galaxies  (of the  FR~I type)  which have  a CCC  do not  have  its UV
counterpart.  These missing  nuclei  seem to  be  associated with  the
presence of extended ($\sim 100$ pc--scale) dusty disks which are seen
almost  edge--on  in these  galaxies,  and  might  absorb the  nuclear
emission in the  UV band.  However, as the nuclei  are clearly seen in
the  optical,  the  amount  of  absorption  must  be  small,  and  not
comparable  to   the  much  higher   column  densities  characterizing
absorbing ``tori'' in  other classes of AGN.  The  high detection rate
of unresolved  nuclei in the  UV among FR~I sources  further indicates
that their nuclei  are generally seen directly and  absorbing tori are
not a common characteristic among such low power radio galaxies.

Of the FR~II, three are  classified as broad--lined radio galaxies and
show the brightest nuclei. The only non--broad lined FR~II which shows
a (fainter) CCC is 3C~388, which is classified as a LEG.

We have shown that by combining the UV data with the already available
optical and radio information we can further investigate the origin of
these nuclei.   In a  plane formed by  the optical--UV  spectral index
versus the optical excess with respect to the radio core emission, CCC
occupy  different  regions.   This  can  be well  explained  if  their
position in such plane is  strictly connected to the emission process.
In particular,  the bright nuclei of broad--lined  FR~II are explained
with thermal  emission from  the accretion disk,  while the  other CCC
(all FR~I and one FR~II,  3C~388) are compatible with being originated
by synchrotron  radiation from the jet.  The  strong variability found
in the case of 3C~317 is a further clue of its synchrotron jet origin.

A  major  result of  this  work  is that  only  a  moderate amount  of
absorption, whose magnitude appears to be linked to the orientation of
the source, is needed to account for the wide range of $\alpha_{o,UV}$
spanned by FR~I nuclei.  Extinction  can be higher than $A_V \sim 1-2$
only in highly misoriented  galaxies which clearly show extended dusty
structures.   This indeed  constitutes  the first  direct evidence  of
nuclear absorption in FR~I  radio galaxies.  Although supported by the
presence of only one object, it appears that 3C~388 belongs to a class
of FR~II with FR~I--like  nuclei, in agreement with previous findings.
In the framework  of the unification models, these  objects might well
represent  the  parent population  of  BL  Lacs  with an  FR~II  radio
morphology and power \cite{kollgaard96,cassaro}.

FR~II in which no nuclear source  is seen are expected in the frame of
the unification  models, and  we have  found 11 FR~II  with no  UV CCC
which might be the obscured counterparts of BLRG. However, in order to
firmly establish their role  a detailed comparison of other properties
of these sources (e.g.   emission lines, X-ray spectra, extended power
and morphology) with  those of the unabsorbed FR~II  has to be carried
out.  If they indeed harbor an obscured quasar nucleus, these galaxies
should also  show strong IR nuclear  components, as it  has been found
(at  10   $\mu$m)  for  3C~405   \cite{whysong}.   Unfortunately,  the
incompleteness of  the sample prevents us from  drawing any conclusion
on the geometry and covering factor of the obscuring material based on
the relative number counts of obscured an unobscured objects.

These results  provide further support  for the idea that  the nuclear
structure of FR~I is different  from other AGN, in which signatures of
the presence  of nuclear optically  thick dusty tori are  often found.
In these  low power radio  galaxies, the absorbing material  cannot be
distributed  either as a  spherical structure  nor as  a thin  disk in
which  the absorbing material  is well  confined.  The  extended dusty
disks  often observed  in such  galaxies  can well  account for  their
observed  properties.   Having  assessed  the nature  of  the  nuclear
emission, the  study of the  possible connection between  the extended
dusty structures, the feeding mechanism  of the central black hole and
the nature of the accretion  process in the different classes of radio
galaxies is a promising future perspective for this work.

\acknowledgments

This  work has  been partially  supported  by the  STScI (DDRF)  grant
D0001.82258.    This  research   has   made  use   of  the   NASA/IPAC
Extragalactic Database  (NED) which is operated by  the Jet Propulsion
Laboratory,  California Institute of  Technology, under  contract with
the National Aeronautics and Space Administration.
The authors wish to thank the anonymous referee for useful  comments and
suggestions that improved the paper.




\clearpage

\clearpage

\clearpage

\end{document}

%% file: tab1.tex
\begin{deluxetable}{l l c c c c c}
\tablewidth{0pt}
\tablecaption{The sample}
\tablehead{
\colhead{Name}  & \colhead{Morph. Class.}     &
\colhead{Spectral Class.} & \colhead{redshift}  & 
\colhead{$F_r$} &
\colhead{Opt CCC} & \colhead{UV CCC} \\
\colhead{~ }   & \colhead{~ }   &\colhead{~ }   & \colhead{$z$} &\colhead{mJy}   & 
\colhead{~ } & 
\colhead{~ } }
\startdata
3C~29     &  FR~I     &  --   & 0.0448      &  93.0      &  YES   &  YES    \\ 
3C~35     &  FR~II    &  LEG  & 0.0670      &  22.68     &  NO    &  NO     \\        
3C~40     &  FR~II    &  LEG  & 0.0180      &  626.9     &  NO    &  NO     \\ 
3C~78     &  FR~I     &       & 0.029       &  964       &  YES   & YES    \\
3C~66B    &  FR~I     &  --   & 0.0215      &  182.0     &  YES   &  YES    \\ 
3C~192    &  FR~II    &  HEG  & 0.0600      &   8.1      &  NO    &  NO     \\ 
3C~198    &  FR~II    &  HEG  & 0.0820      &   --       &  YES   &  YES    \\ 
3C~227    &  FR~II    & BLRG  & 0.0860      &   23.23    &  YES   &  YES    \\ 
3C~236    &  FR~II    &  LEG  & 0.0990      &  190       &  NO    &  NO     \\ 
3C~264    &  FR~I     &  --   & 0.0206      &  200.0     &  YES   &  YES    \\  
3C~270    &  FR~I     &  --   & 0.0074      &  308.0     &  YES   &  NO     \\ 
3C~274 (M~87) &  FR~I &  --   & 0.0037      &  4000.0    &  YES   &  YES    \\   
3C~285    &  FR~II    &  HEG  & 0.0790      &  7.7       &  YES    &  NO     \\ 
3C~293    &  FR~II    &  LEG  & 0.0452      &  100.0     &  NO    &  NO     \\ 
3C~296    &  FR~I     &  --   & 0.0237      &  77.0      &  YES   &  NO     \\ 
3C~305    &  FR~I     &  --   & 0.0410      &  29.5      &  NO    &  NO     \\ 
3C~310    &  FR~I     &  --   & 0.0540      &  80.0      &  YES   &  YES    \\  
3C~317    &  FR~I     &  --   & 0.0342      &  391.0     &  YES   &  YES    \\ 
3C~321    &  FR~II    &  HEG  & 0.0960      &  37.5      &  NO    &  NO     \\ 
3C~326    &  FR~II    &  LEG  & 0.0890      &  15.7      &  NO    &  NO     \\ 
3C~338    &  FR~I     &  --   & 0.0303      &  105.0     &  YES   &  YES    \\  
3C~353    &  FR~II    &  LEG  & 0.0300      &  203.5     &  NO    &  NO     \\ 
3C~382    &  FR~II    & BLRG  & 0.0580      &  217.4     &  YES   &  YES    \\   
3C~388    &  FR~II    &  LEG  & 0.0910      &  75.67     &  YES   &  YES    \\  
3C~390.3  &  FR~II    & BLRG  & 0.0560      &   414      &  YES   &  YES    \\       
3C~449    &  FR~I     &  --   & 0.0181      &  37.0      &  YES   &  YES    \\ 
3C~452    &  FR~II    &  HEG  & 0.0810      &  150       &  NO    &  NO     \\ 
3C~465    &  FR~I     &  --   & 0.0301      &  270.0     &  YES   &  YES    \\ 
\enddata

\medskip

Classifications and  data from  the literature for  the sample  of 3CR
radiogalaxies.   (1) 3C name  of the  source; (2)  radio morphological
classification;  (3) optical  spectral classification,  as  taken from
Jackson \&  Rawlings (1997); (4)  Redshift (from NED);
(5) radio core flux at 5 GHz collected from the literature (Giovannini
et al. 1988,  Zirbel \& Baum 1995); (6) and  (7) presence/absence of a
CCC in the  HST images (from Paper~I, Paper~II, Allen  et al. 2001 and
this work).
\label{tab1}

\end{deluxetable}



%% file: tab2.tex
\begin{deluxetable}{l c c c c c c}
\tablewidth{0pt}
\tablecaption{Log of HST optical and UV observations}
\label{log}
\tablehead{
\colhead{Name}  & \colhead{Instrument}     & \colhead{Filter}     & \colhead{Obs. date}     &
\colhead{Instrument} & \colhead{Filter}  & 
\colhead{Obs. Date}  }
\startdata
3C~29   &  WFPC2         &   F702W    & Jan 12 1995  & STIS NUV--MAMA  & F25SRF2   &   Jun 08 2000     \\     
3C~35   &  WFPC2         &   F702W    & Mar 12 1994  & STIS NUV--MAMA  & F25SRF2   &   Oct 10 1999     \\            
3C~40   &  WFPC2         &   F702W    & Jul 18 1994  & STIS NUV--MAMA  & F25SRF2   &   Jun 06 2000     \\     
3C~66B  &  WFPC2         &   F814W    & Jan 31 1999  & STIS NUV--MAMA  & F25SRF2   &   Jul 13 2000     \\    
3C~78   &  STIS CCD      &   F28X50LP & Mar 15 2000  & STIS NUV--MAMA  & F25QTZ    &   Mar 15 2000     \\     
3C~192  &  WFPC2         &   F555W    & Jan 13 1997  & STIS NUV--MAMA  & F25SRF2   &   Mar 23 2000     \\     
3C~198  &  WFPC2         &   F702W    & Mar 20 1994  & STIS NUV--MAMA  & F25SRF2   &   Apr 23 2000     \\     
3C~227  &  WFPC2         &   F702W    & May 19 1995  & STIS NUV--MAMA  & F25SRF2   &   Jan 25 2000     \\     
3C~236  &  WFPC2         &   F702W    & Oct 19 1994  & STIS NUV--MAMA  & F25SRF2   &   Oct 05 1999     \\     
3C~264  &  STIS CCD      &   F28X50LP & Feb 13 2000  & STIS NUV--MAMA  & F25CN182  &   Feb 12 2000     \\      
3C~270  &  WFPC2         &   F791W    & Dec 13 1994  & STIS NUV--MAMA  & F25SRF2   &   Mar 05 2000     \\     
3C~274  &  WFPC2         &   F814W    & May 11 1999  & STIS NUV--MAMA  & F25QTZ    &   May 17 1999     \\       
3C~285  &  WFPC2         &   F702W    & Feb 05 1994  & STIS NUV--MAMA  & F25SRF2   &   Apr 16 2000     \\     
3C~293  &  WFPC2         &   F702W    & Jan 15 1995  & STIS NUV--MAMA  & F25SRF2   &   Jun 14 2000     \\     
3C~296  &  WFPC2         &   F702W    & Dec 14 1994  & STIS NUV--MAMA  & F25SRF2   &   Apr 15 2000     \\
3C~305  &  WFPC2         &   F702W    & Sep 04 1994  & STIS NUV--MAMA  & F25SRF2   &   Apr 27 2000     \\
3C~310  &  WFPC2         &   F702W    & Sep 12 1994  & STIS NUV--MAMA  & F25SRF2   &   Jun 10 2000     \\
3C~317  &  FOC           &   F555W    & Mar 05 1994  & FOC             & F210M     &   Mar 05 1994     \\
3C~321  &  WFPC2         &   F702W    & Apr 29 1995  & STIS NUV--MAMA  & F25SRF2   &   Jun 05 2000     \\
3C~326  &  WFPC2         &   F702W    & Apr 29 1995  & STIS NUV--MAMA  & F25SRF2   &   Mar 12 2000     \\
3C~338  &  WFPC2         &   F702W    & Sep 09 1994  & STIS NUV--MAMA  & F25SRF2   &   Jun 04 2000     \\
3C~353  &  WFPC2         &   F702W    & Mar 18 1995  & STIS NUV--MAMA  & F25SRF2   &   Jun 22 2000     \\
3C~382  &  WFPC2         &   F702W    & Jun 25 1994  & STIS NUV--MAMA  & F25CN182  &   Feb 23 2000     \\
3C~388  &  WFPC2         &   F702W    & Sep 18 1994  & STIS NUV--MAMA  & F25SRF2   &   Jun 02 2000     \\
3C~390. &  WFPC2         &   F702W    & Sep 20 1994  & STIS NUV--MAMA  & F25CN182  &   Aug 10 1999     \\
3C~449  &  WFPC2         &   F702W    & Aug 06 1994  & STIS NUV--MAMA  & F25SRF2   &   Apr 16 2000     \\ 
3C~452  &  WFPC2         &   F702W    & May 05 1994  & STIS NUV--MAMA  & F25SRF2   &   Jan 30 2000     \\
3C~465  &  WFPC2         &   F814W    & Jul 03 2000  & STIS NUV--MAMA  & F25SRF2   &   May 25 2000     \\
\enddata
\end{deluxetable}



%% file: tab3.tex
\begin{deluxetable}{l c c c c}
\tablewidth{0pt}
\tablecaption{Observed fluxes of optical and UV CCC}
\label{cccfluxes}
\tablehead{
\colhead{Source name}  & \colhead{$F_{opt}$}     &
\colhead{$\lambda_{opt}$} & \colhead{$F_{UV}$}  & 
\colhead{$\lambda_{UV}$}  \\
\colhead{~ }   &   \colhead{erg cm$^{-2}$ s$^{-1}$ \AA$^{-1}$} & 
\colhead{\AA} &  \colhead{erg cm$^{-2}$ s$^{-1}$ \AA$^{-1}$} &
\colhead{\AA} }
\startdata
FR~I      &            ~               &    ~       &            ~                  &      ~      \\
\tableline
3C~29     &   $5.8 \times 10^{-18}$   &   7000    &    $ 2.1 \times 10^{-19}$    &    2528    \\ 
3C~66B    &   $2.7 \times 10^{-17}$   &   8086    &    $ 4.5 \times 10^{-17}$    &    2528    \\ 
3C~78     &   $3.8 \times 10^{-16}$   &   7216    &    $ 2.8 \times 10^{-16}$    &    2475    \\
3C~264    &   $1.6 \times 10^{-16}$   &   7216    &    $ 3.0 \times 10^{-16}$    &    2078    \\  
3C~270$^a$&   $5.1 \times 10^{-18}$   &   7930    &    $<1.0 \times 10^{-18}$    &    2528    \\ 
3C~274    &   $3.4 \times 10^{-16}$   &   8086    &    $ 3.8 \times 10^{-16}$    &    2475    \\   
3C~310    &   $3.5 \times 10^{-18}$   &   7000    &    $ 1.6 \times 10^{-18}$    &    2528    \\  
3C~317    &   $2.0 \times 10^{-17}$   &   5508    &    $ 4.7 \times 10^{-18}$    &    2213    \\  
3C~338    &   $1.0 \times 10^{-17}$   &   7000    &     $3.9 \times 10^{-18}$    &    2528    \\  
3C~449    &   $1.8 \times 10^{-17}$   &   7000    &     $1.9 \times 10^{-19}$    &    3256    \\  
3C~465    &   $1.0 \times 10^{-17}$   &   8086    &     $2.1 \times 10^{-18}$    &    2528    \\ 
\tableline
FR~II     &           ~                &    ~       &            ~                  &     ~       \\
\tableline
3C~198    &   $4.9 \times 10^{-17}$   &   7000    &    $ 1.1 \times 10^{-16}$    &    2528    \\ 
3C~227    &   $2.9 \times 10^{-16}$   &   7000    &    $ 5.9 \times 10^{-16}$    &    2528    \\ 
3C~382    &   $4.9 \times 10^{-15}$   &   7000    &     $1.6 \times 10^{-14}$    &    2078    \\   
3C~388    &   $6.7 \times 10^{-18}$   &   7000    &     $2.0 \times 10^{-18}$    &    2528    \\  
3C~390.3  &   $1.1 \times 10^{-15}$   &   7000    &     $2.0 \times 10^{-15}$    &    2078    \\       
\enddata

\medskip

$^a$ For 3C~270 an upper limit to the UV CCC has been derived (see section 3).

\end{deluxetable}

%% file: tab4.tex
\begin{deluxetable}{l c c c c c c c c}
\tablewidth{0pt}
\tablecaption{Dereddened UV and optical data of the CCC}
\tablehead{
\colhead{Source name}  & \colhead{Spectr. Class.} & 
\colhead{$\log F_{opt}$} & 
\colhead{$\log F_{UV}$}  & 
\colhead{$\log L_{opt}$} & 
\colhead{$\log L_{UV}$}  &
\colhead{$\alpha_{\rm o,UV}$} &
\colhead{$\sigma_\alpha$} &
\colhead{E(B-V)}}  
\startdata
FR~I      &      &             &                &         &              &        &          &            \\
\tableline        
3C~29     &      & -27.99      &    -30.25      &  26.57  &    24.31     &   5.1  &    0.2   &   0.036    \\ 
3C~66B    &      & -27.18      &    -27.80      &  26.76  &    26.13     &   1.2  &    0.1   &   0.080    \\ 
3C~78     &      & -26.03      &    -26.74      &  28.15  &    27.44     &   1.5  &    0.1   &   0.173    \\
3C~264    &      & -26.54      &    -27.28      &  27.36  &    26.61     &   1.4  &    0.1   &   0.023    \\  
3C~270    &      & -27.96      &    -29.62      &  25.05  &    23.39     &  $>$ 3.4  &    -- &   0.018    \\ 
3C~274    &      & -26.12      &    -27.05      &  26.30  &    25.37     &   1.8  &    0.1   &   0.022    \\   
3C~310    &      & -28.21      &    -29.35      &  26.50  &    25.36     &   2.6  &    0.1   &   0.042    \\  
3C~317    &      & -27.65      &    -28.97      &  26.68  &    25.35     &   3.3  &    0.2   &   0.037    \\  
3C~338    &      & -27.78      &    -29.05      &  26.44  &    25.17     &   2.9  &    0.1   &   0.012    \\  
3C~449    &      & -27.38      &    -29.83      &  26.40  &    23.95     &   7.3  &    0.2   &   0.167    \\  
3C~465    &      & -27.61      &    -29.16      &  26.60  &    25.06     &   3.1  &    0.1   &   0.069    \\ 
\tableline        
FR~II     &      &             &                &         &              &        &          &            \\
\tableline        
3C~198    &  HEG    & -27.08      &    -27.56      &  27.98  &    27.50     &   1.1  &    0.1   &   0.026    \\ 
3C~227    & BLRG    & -26.30      &    -26.83      &  28.79  &    28.26     &   1.2  &    0.1   &   0.026    \\ 
3C~382    & BLRG    & -25.03      &    -25.37      &  29.74  &    29.40     &   0.6  &    0.1   &   0.070    \\   
3C~388    &  LEG    & -27.89      &    -29.15      &  27.25  &    25.99     &   2.8  &    0.2   &   0.080    \\  
3C~390.3  & BLRG    & -25.68      &    -26.27      &  29.06  &    28.47     &   1.1  &    0.1   &   0.071    \\       
\enddata

\medskip
Fluxes (in units of erg cm$^{-2}$ s$^{-1}$ Hz$^{-1}$) and luminosities
(erg  s$^{-1}$  Hz$^{-1}$)  dereddened  taking into  account  galactic
absorption (column (8), from NED).

\label{ccc2}

\end{deluxetable}